\documentclass[twocolumn,prd,amssymb,aps,nofootinbib,showpacs]{revtex4}
\usepackage[usenames,dvipsnames]{color}
\usepackage{ulem}
\usepackage{graphicx}
\usepackage{amssymb}
\usepackage{amsmath}
\usepackage{epstopdf}
\usepackage{aas_macros}
\newcommand{\bbe}{\begin{equation}}
\newcommand{\be}{\begin{equation}}
\newcommand{\ee}{\end{equation}}
\newcommand{\bea}{\begin{eqnarray}}
\newcommand{\eea}{\end{eqnarray}}
\newcommand{\beaa}{\begin{eqnarray*}}
\newcommand{\eeaa}{\end{eqnarray*}}
\newcommand{\ben}{\begin{enumerate}}
\newcommand{\een}{\end{enumerate}}

\begin{document}
\title{Tidal deformability of neutron stars with realistic equations of
state and their gravitational wave signatures in binary inspiral}
\author{Tanja Hinderer$^1$, Benjamin D. Lackey$^2$ , Ryan N.
Lang$^{3,4}$, Jocelyn S. Read$^5$}

\affiliation{ $^1$Theoretical Astrophysics, California Institute of Technology, Pasadena, CA 91125, USA\\
$^2$Department of Physics, University of Wisconsin--Milwaukee, P.O. Box 413, Milwaukee, WI 53201, USA\\
$^3$Department of Physics and MIT Kavli Institute, MIT, 77 Massachusetts Avenue, Cambridge, MA 02139, USA\\
$^4$Gravitational Astrophysics Laboratory, NASA Goddard Space Flight Center, 8800 Greenbelt Rd., Greenbelt, MD 20771, USA\\
$^5$Max-Planck-Institut f\"{u}r Gravitationsphysik
Albert-Einstein-Institut, Am M\"{u}hlenberg 1, 14476 Potsdam, Germany}

\begin{abstract}
  The early part of the gravitational wave signal of binary
  neutron star inspirals can potentially yield robust information on
  the nuclear equation of state. The influence of a star's internal
  structure on the waveform is characterized by a single parameter:
  the tidal deformability $\lambda$, which measures the star's
  quadrupole deformation in response to the companion's perturbing
  tidal field.  We calculate $\lambda$ for a wide range of equations of
  state and find that the value of $\lambda$ spans an order of magnitude for the range of equation
   of state models considered.

An analysis of the feasibility of discriminating between neutron star equations of state
with gravitational wave observations of the early part of the
inspiral reveals that the measurement error in $\lambda$ increases
steeply with the total mass of the binary. Comparing the errors with
the expected range of $\lambda$, we find that Advanced LIGO
observations of binaries at a distance of 100~Mpc will probe only unusually stiff equations of state,
while the proposed Einstein Telescope is likely to see a clean tidal
signature.
\end{abstract}

\pacs{
04.40.Dg, 
26.60.Kp, 
97.60.Jd,  
95.85.Sz 
}

\maketitle

\section{Introduction and summary}

The observation of inspiraling binary neutron stars (NSs) with
ground-based gravitational-wave detectors such as LIGO and Virgo may
provide significantly more information about neutron-star structure,
and the highly uncertain equation of state (EOS) of neutron-star
matter, than is currently available. The available electromagnetic
observations of neutron stars provide weak constraints from properties
such as the star's mass, spin, and gravitational
redshift (see for example~\cite{LattimerPrakash2007,
ReadLackeyOwenFriedman2009}).  Simultaneous measurements of both the mass and
radius of a neutron star~\cite{Ozel2006, LeahyMorsinkCadeau2008, OzelGuverPsaltis2008,
GuverOzeletal2008,Leahyet2009}, on the other hand, have the potential to make
significantly
stronger constraints.  These measurements, however, depend on detailed modeling of the radiation mechanisms at the neutron-star surface and absorption in the interstellar medium, and are subject to systematic uncertainties.

Another possibility for obtaining information about the neutron star EOS is
from the inspiral of binary neutron stars due to gravitational radiation.
The tidal distortion of neutron stars in a binary system links the EOS
describing neutron-star matter to the gravitational-wave emission during
the inspiral. Initial estimates showed that for LIGO, tidal effects
change the phase evolution only at the end of inspiral, and that point
particle post-Newtonian waveforms can be used for template-based
detection~\cite{Kochanek1992, BildstenCutler1992, LaiWiseman}. With
the projected sensitivities of later-generation detectors, however, effects
which can be neglected for the purpose of detection may become measurable
in the strongest observed signals.

While EOS effects are largest during the late inspiral and merger
of two neutron stars where numerical simulations must
be used to predict the signal, Flanagan and Hinderer showed that a
small but clean tidal signature arises in the inspiral below
400~Hz~\cite{FlanaganHinderer2008}. This signature amounts to a
phase correction which can be described in terms of a single
EOS-dependent tidal deformability parameter $\lambda$, namely the
ratio of each star's induced quadrupole to the tidal field
of its companion. The parameter $\lambda$
depends on the EOS via both the
NS radius $R$ and a
dimensionless quantity $k_2$, called the Love
number~\cite{BrookerOlle1955, MoraWill2004, BertiIyerWill2008}: $\lambda= 2/(3G) k_2 R^5$.

The relativistic Love numbers of
polytropic\footnote{Polytropes are often written in two forms.
The first form is expressed as $p=K\epsilon^{1+1/n}$, where $p$ is
the pressure, $\epsilon$ is the energy density, $K$ is a pressure constant,
 and $n$ is the polytropic index.  The second form, is given by $p=K\rho^{1+1/n}$,
 where $\rho$ is the rest-mass density, defined as the baryon number density times
  the baryon rest mass.  The first form was mainly used in the recent
  papers~\cite{Hinderer2008, DamourNagar2009, BinningtonPoisson2009}.
  However, the second form is more commonly used in the neutron-star literature
  and is more closely tied to the thermodynamics of a Fermi gas.
  We will use both forms as was done in Ref.~\cite{DamourNagar2009}.} EOS
  were examined first by Flanagan and Hinderer~\cite{FlanaganHinderer2008, Hinderer2008}
  and later by others in more detail~\cite{DamourNagar2009, BinningtonPoisson2009}.
  Flanagan and Hinderer also examined the measurability of the tidal deformability
    of polytropes and suggested that Advanced LIGO could start to place
interesting constraints on $\lambda$ for nearby events.
However, they used incorrect values for $k_2$,
which overestimated $\lambda$ by a factor of $\sim 2-3$ and were
therefore overly optimistic about the potential measurability. In
addition, polytropes are known to be a poor approximation to the
neutron star equation of state, and there may be significant differences in
the tidal deformability between polytropes and ``realistic'' EOS. In this
paper, we calculate the deformability for realistic EOS, and show
that a tidal signature is actually only marginally detectable with
Advanced LIGO. 

In Sec.~\ref{sec:calculation} we
describe how the Love number and tidal deformability can be
calculated for tabulated EOS.  We
use the equations for $k_2$
developed in \cite{Hinderer2008}, which arise from a linear
perturbation of the Oppenheimer-Volkoff (OV) equations of
hydrostatic equilibrium.  In Sec.~\ref{sec:candidateeos} we then calculate $k_2$ and $\lambda$ as a
function of mass for several EOS commonly found in the literature.
We find that, in contrast to the Love number, the tidal
deformability has a wide range of values, spanning roughly an order
of magnitude over the observed mass range of neutron stars in binary
systems.

As discussed above, the direct practical importance of the stars' tidal
deformability for gravitational wave observations of NS
binary inspirals is that it encodes the EOS influence on the
waveform's phase evolution during the early portion of the
signal, where it is accurately modeled by post-Newtonian (PN) methods.  In this regime, the influence of tidal effects is
only a small correction to the point-mass dynamics.  However,
 when the signal is integrated
against a theoretical waveform template over
many cycles, even a small contribution to the phase
evolution can be detected and could give information about the NS
structure.

Following~\cite{FlanaganHinderer2008}, we calculate in
Sec.~\ref{sec:radiation} the measurability of the tidal
deformability for a wide range of equal- and unequal- mass binaries,
covering the entire expected range of NS masses and EOS, and with
proposed detector sensitivity curves for second- and third-
generation detectors. We show that the measurability of $\lambda$ is
quite sensitive to the total mass of the system, with very low-mass
neutron stars contributing significant phase corrections that are
optimistically detectable in Advanced LIGO, while larger-mass
neutron stars are more difficult to distinguish from the $k_2=0$
case of black holes~\cite{DamourNagar2009, BinningtonPoisson2009}.
In third-generation detectors, however, the tenfold increase in
sensitivity allows a finer discrimination between equations of state
leading to potential measurability of a large portion of proposed
EOSs over most of the expected neutron star mass range.

We conclude by briefly considering how the errors could be
improved with a more careful analysis of the detectors and extension of
the understanding of EOS effects to higher frequencies.

Finally, in the Appendix we compute the leading order EOS-dependent
corrections to our model of the tidal effect and derive explicit
expressions for the resulting corrections to the waveform's phase evolution, extending the analysis of
Ref.~\cite{FlanaganHinderer2008}. Estimates of the size of the phase
corrections show that the main source of error are post-1 Newtonian
corrections to the Newtonian tidal effect itself, which are
approximately twice as large as other, EOS-dependent corrections at
a frequency of 450 Hz. Since these point-particle corrections do not
depend on unknown NS physics, they can easily be incorporated into
the analysis. A derivation of the explicit post-Newtonian correction
terms is the subject of Ref.~\cite{Vinesflanagan}.

\textit{Conventions}: We set $G=c=1$.

\section{Calculation of the Love number and tidal deformability}
\label{sec:calculation}
As in \cite{FlanaganHinderer2008} and \cite{Hinderer2008}, we consider
a static, spherically symmetric star, placed in a static external
quadrupolar tidal field $\mathcal{E}_{ij}$. To linear order,
we define the tidal deformability $\lambda$ relating the star's
induced quadrupole moment $Q_{ij}$ to the external tidal field,
\begin{equation}
  Q_{ij} = - \lambda \mathcal{E}_{ij}.\label{eq:lambdadef}
\end{equation}
The coefficient $\lambda$ is related to the $l=2$
dimensionless tidal Love number $k_2$ by
\begin{equation}
k_2 = \frac{3}{2} \lambda R^{-5}.
\end{equation}

The star's quadrupole moment $Q_{ij}$  and the
external tidal field $\mathcal{E}_{ij}$ are
defined to be coefficients in an asymptotic expansion of the total
metric at large distances $r$ from the star. This expansion includes,
for the metric component $g_{tt}$ in asymptotically Cartesian, mass-centered coordinates, the standard gravitational potential $m/r$, plus two
leading order terms arising from the perturbation, one describing an
external tidal field growing with $r^2$ and one describing the
resulting tidal distortion decaying with $r^{-3}$:
\begin{eqnarray}
-\frac{\left(1+g_{tt} \right)}{2} &=& - \frac{m}{r}-\frac{3Q_{ij}}{2 r^3} n^i n^j +\ldots
+\frac{{\cal E}_{ij}}{ 2} r^2 n^i n^j + \ldots, \nonumber\\
\label{eq:metric}
\end{eqnarray}
 where $n^i=x^i/r$ and $Q_{ij}$ and $\mathcal{E}_{ij}$ are both symmetric and traceless.
 The relative size of these
multipole components of the perturbed spacetime gives the constant
$\lambda$ relating the quadrupole deformation to the external tidal
field as in Eq. (\ref{eq:lambdadef}).

To compute the metric \eqref{eq:metric}, we use
the method discussed in \cite{Hinderer2008}. We consider the
problem of a linear static perturbation expanded in spherical
harmonics following \cite{ThorneCampolattaro1967}.  Without loss of
generality we can set the azimuthal number $m=0$, as the tidal
deformation will be axisymmetric around the line connecting the two
stars which we take as the axis for the spherical harmonic
decomposition. Since we will be interested in applications to the
early stage of binary inspirals, we will also specialize to the
leading order for tidal effects, $l=2$.

Introducing a linear $l=2$ perturbation onto the spherically
symmetric star results in a static (zero-frequency), even-parity
perturbation of the metric, which in the Regge-Wheeler
gauge~\cite{ReggeWheeler} can be simplified~\cite{Hinderer2008} to
give
\begin{eqnarray}
ds^2 &=& - e^{2\Phi(r)} \left[1 + H(r)
Y_{20}(\theta,\varphi)\right]dt^2
\nonumber\\
& & + e^{2\Lambda(r)} \left[1 - H(r)
Y_{20}(\theta,\varphi)\right]dr^2
\nonumber \\
& & + r^2 \left[1-K(r) Y_{20}(\theta,\varphi)\right] \left( d\theta^2+ \sin^2\theta
d\varphi^2 \right),
\nonumber\\
& &
\end{eqnarray}
where $K(r)$ is related to $H(r)$ by $K'(r)=H'(r)+2 H(r) \Phi'(r)$.  Here primes denote derivatives with respect to $r$.
The corresponding perturbations of the perfect
fluid stress-energy tensor components are $\delta
T_0^{\,0}=-\delta\epsilon(r) Y_{20} (\theta,\varphi)$ and $\delta
T_i^{\,i}=\delta p(r) Y_{20}(\theta,\varphi)$, where $\epsilon$ is
the energy density and $p$ the pressure. The function $H(r)$
satisfies the differential equation
\begin{eqnarray}
\left(-\frac{6e^{2\Lambda}}{r^2}-2(\Phi ')^2+2\Phi
  ''+\frac{3}{r}\Lambda '\right.\nonumber\\
\left.+\frac{7}{r}\Phi '- 2 \Phi'
  \Lambda '+ \frac{f}{r}(\Phi '+\Lambda ')\right)H\nonumber\\
+ \left(\frac{2}{r}+\Phi '-\Lambda'\right)H' + H'' =0
.\label{eq:h}
\end{eqnarray}
Here $f$ is given by
\begin{equation}
\delta\epsilon=f \delta p
\end{equation}
which for slow changes in matter configurations corresponds to $f
=d\epsilon/d p$.

The method of calculating the
tidal perturbation for a general equation of state
table is similar to the method of calculating moment of inertia in
the slow rotation approximation \cite{HartleThorne1968}. The
specific implementation we use follows the moment of inertia
calculation in Appendix A of \cite{ReadLackeyOwenFriedman2009}, via
an augmentation of the OV system of equations\footnote{Here we
present the equations
  in terms of the radial coordinate $r$; the extension to the enthalpy
  variable $\eta$ used in \cite{ReadLackeyOwenFriedman2009} is
  straightforward.}:
\begin{eqnarray}
e^{2 \Lambda} &=& \left(1-\frac{2m_r}{r}\right)^{-1},\label{eq:lambda}\\
\frac{d\Phi}{dr} &=& - \frac{1}{\epsilon+p}\frac{dp}{dr}, \label{eq:dphidr}\\
\frac{dp}{dr} &=& -(\epsilon+p)\frac{m_r+4\pi r^3 p}{r(r-2m_r)}, \label{eq:dpdr}\\
\frac{dm_r}{dr} &=& 4 \pi r^2 \epsilon.  \label{eq:dmdr}
\end{eqnarray}
The second-order differential equation for $H$ is separated
into a first-order system of ODEs
in terms of the usual OV quantities $m_r$\footnote{We use $m_r$ for the mass enclosed within radius $r$ instead of $m(r)$ to avoid confusion with the total mass of the star, which we will label $m$.}, $p(r)$, and $\epsilon(p)$, as well as the additional functions $H(r)$, $\beta(r) =
dH/dr$, and the equation of state function $f(p)$ (recall $f =
d\epsilon/dp$):
\bea \frac{dH}{dr}&=& \beta\\
\frac{d\beta}{dr}&=&2 \left(1 - 2\frac{m_r}{r}\right)^{-1} H\left\{-2\pi
  \left[5\epsilon+9
    p+f(\epsilon+p)\right]\phantom{\frac{3}{r^2}} \right. \nonumber\\
&& \quad \left. +\frac{3}{r^2}+2\left(1 - 2\frac{m_r}{r}\right)^{-1}
  \left(\frac{m_r}{r^2}+4\pi r p\right)^2\right\}\nonumber\\
&&+\frac{2\beta}{r}\left(1 -
  2\frac{m_r}{r}\right)^{-1}\left\{-1+\frac{m_r}{r}+2\pi r^2
  (\epsilon-p)\right\}.\nonumber\\\eea
These are combined with
Eqs.~(\ref{eq:lambda})--(\ref{eq:dmdr}), and the augmented system is
solved simultaneously. The system is integrated outward starting
just outside the center using the expansions $H(r)=a_0 r^2$ and
$\beta(r)=2a_0r$ as $r \to 0$.  The constant
$a_0$ determines how much the star is deformed and can be chosen
arbitrarily as it cancels in the expression for the Love number.
The ODE for $H(r)$ outside the star, where $T_{\mu\nu}=0$,
has a general solution in terms of associated Legendre functions
$Q_2^2\left(r/m-1\right) \sim r^{-3}$ at large $r$, and
$P_2^2\left(r/m-1\right) \sim r^{2}$ at large $r$.  The boundary conditions
that determine the unique choice of this solution follow from matching
the interior and exterior solutions and their first derivatives at
the boundary of the star, where $r=R$.  By comparison with Eq.~\eqref{eq:metric}, the coefficients of the
 external solution can then be identified
with the axisymmetric tidal field and quadrupole
moment via $\mathcal{E} Y_{20}(\theta, \varphi) = \mathcal{E}_{ij}n^in^j$, and
 $Q Y_{20}(\theta, \varphi) = Q_{ij}n^in^j = -\lambda\mathcal{E}_{ij}n^in^j$
 as was done in~\cite{Hinderer2008}.  Here, $\mathcal{E}$
 and $Q$ are the magnitudes
 of the $l=2,~ m=0$ spherical harmonic coefficients of
 the tidal tensor and quadrupole moment respectively.

Defining the quantity
\begin{equation}
y = \frac{ R\, \beta(R)} {H(R)}
\end{equation}
for the internal solution, the $l=2$ Love number is

\begin{eqnarray}
k_2 &=& \frac{8C^5}{5}(1-2C)^2[2+2C(y-1)-y]\nonumber\\
      & & \times\bigg\{2C[6-3y+3C(5y-8)]\nonumber\\
      & & ~~~+4C^3[13-11y+C(3y-2)+2C^2(1+y)]\nonumber\\
      & & ~~~+3(1-2C)^2[2-y+2C(y-1)] \ln(1-2C)\bigg\}^{-1},\nonumber\\
\label{eq:k2}
\end{eqnarray}
where $C=m/R$ is the compactness of the star.

For stars with a nonzero density at the surface (for example strange
quark matter or an incompressible $n=0$ polytrope), the term
$(f/r)(\Phi'+\Lambda')$ in Eq.~(\ref{eq:h}) blows up at the surface $r=R$
and $H'(r)$ is no longer continuous across the surface.  Following the
discussion in~\cite{DamourNagar} for an $n=0$ polytrope, this
discontinuity leads to an extra term in the expression above for $y$:
\begin{equation}
y = \frac{ R\, \beta(R)} {H(R)} - \frac{4\pi R^3 \epsilon_-}{m}
\end{equation}
where $\epsilon_-$ is the density just inside the surface.

\section{Love numbers and tidal deformabilities for candidate EOS}
\label{sec:candidateeos}

Differences between candidate EOS can have a significant effect on
the tidal interactions of neutron stars. In this paper we consider a
sample of EOS from Refs.~\cite{LattimerPrakash2001,
ReadLackeyOwenFriedman2009} with a variety of generation methods and
particle species.  The sample is chosen to include EOS with the
largest range of behaviors for $k_2(m/R)$, $k_2(m)$ and $\lambda(m)$
rather than to fairly represent the different generation methods. We
also restrict ourselves to stars with a maximum mass greater than
1.5~$M_\odot$, which is conservatively low given recent neutron-star
mass observations~\cite{RansomEt2005,
  FreireEt2008, ChampionEt2008, VerbiestEt2008, FreireEt2008b}.  We
consider 7 EOS with just normal $npe\mu$ matter (SLY~\cite{sly4},
AP1 and AP3~\cite{apr}, FPS~\cite{fp}, MPA1~\cite{mpa1}, MS1 and
MS2~\cite{ms}), 8 EOS that also incorporate some combination of
hyperons, pion condensates, and quarks (PS~\cite{ps},
BGN1H1~\cite{balbn1h1}, GNH3~\cite{glendnh3}, H1 and
H4~\cite{LackeyNayyarOwen}, PCL2~\cite{pcl}, ALF1 and
ALF2~\cite{AlfordBraby2005hybrid}), and 3 self-bound strange quark
matter EOS (SQM1-3~\cite{sqm}).  A brief description of these EOS and their
properties can be found in \cite{LattimerPrakash2001,
ReadLackeyOwenFriedman2009}.

\begin{figure}[!htb]
\begin{center}
\includegraphics[width=70mm]{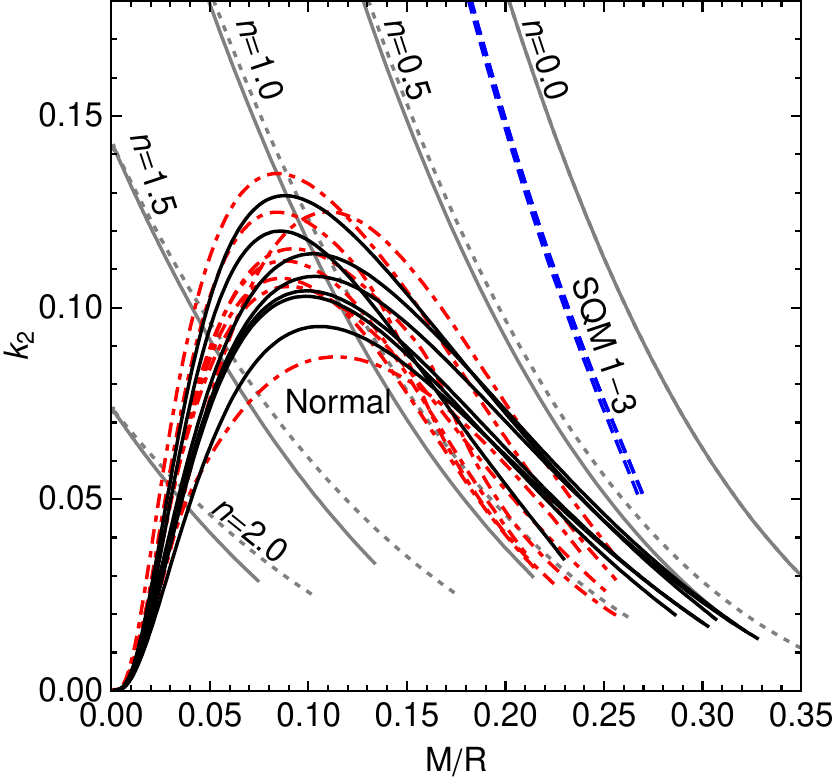}\\
\includegraphics[width=70mm]{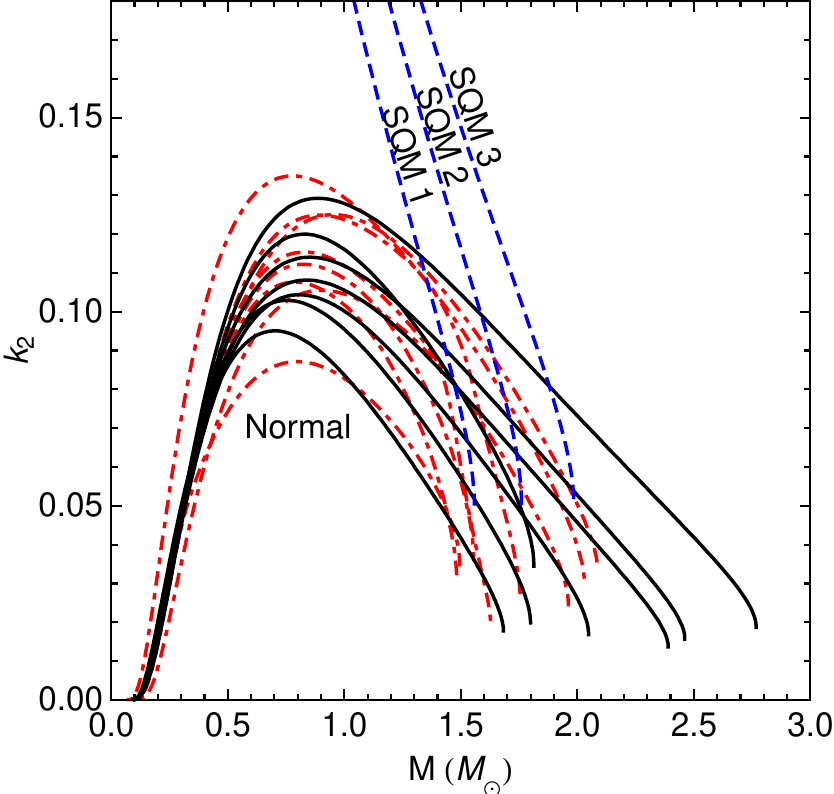}
\end{center}
\caption{ \label{fig:k2} Top panel: Love number as a function of
compactness.  Gray dotted curves are energy density polytropes
($p=K\epsilon^{1+1/n}$), and gray solid curves are rest-mass density
polytropes ($p=K\rho^{1+1/n}$).  Both polytropes are the same for $n=0$.
EOS with only $npe\mu$ matter are solid and those that also incorporate
$\pi$/hyperon/quark matter are dot-dashed.  The three SQM EOS are dashed
and overlap.  They approach the $n=0$ curve at low compactness, where $k_2$
has a maximum value of 0.75 as $m/R \to 0$.  Bottom panel: Love number as a
function of mass for the same set of realistic EOS.  Note that there is
more variation in $k_2$ between different EOS for fixed mass than for fixed
compactness.}
\end{figure}

The generic behavior of the Love number $k_2$ is shown in
the top panel of Fig.~\ref{fig:k2} as a function of compactness $m/R$ for different types of EOS.  The two types of
polytropes, energy and rest-mass density
polytropes, are shown in gray.  They coincide in the 
limit $m/R \to 0$ where $\epsilon \to \rho$ as the star's density
goes to zero, and in the limit $n \to 0$ where $\epsilon(p)$ and $\rho(p)$
are both constant.
This can be seen from the first law of thermodynamics,
\begin{equation}
d\frac{\epsilon}{\rho} = -pd\frac{1}{\rho},
\end{equation}
which relates $\epsilon$ to $\rho$.  

The sequences labeled ``Normal'' correspond to the 15 EOS with a
standard nuclear matter crust, and the 3 sequences labeled ``SQM''
correspond to the crustless EOS SQM1-3 where the pressure is zero
below a few times nuclear density.
Within these two classes, there is little variation
in behavior, so we do not explicitly label each candidate EOS.

The bottom panel of Fig.~\ref{fig:k2} shows $k_2(m)$ for the
 realistic EOS, which is more astrophysically relevant because mass, not
  compactness, is the measurable quantity during binary inspiral.
  Unlike the quantity $k_2(m/R)$, $k_2(m)$ depends on the constant
   $K$ for polytropes, so polytropic EOS are not shown.
   There is more variation in $k_2$ for fixed mass than for fixed compactness.

The behavior of these curves can be understood as follows: The Love
number $k_2$ measures how easily the bulk of the matter in a star is
deformed.  If most of the star's mass is concentrated at the center
(centrally condensed), the tidal deformation will be smaller.  For
polytropes, matter with a higher polytropic index $n$ is softer and
more compressible, so these polytropes are more centrally condensed.  As a result,
$k_2$ decreases as $n$ increases.  The limiting case $n=0$
represents a uniform density star and has the largest Love number
possible.  The Love number also decreases with increasing
compactness, and from Eq.~\eqref{eq:k2} it can be seen that $k_2$
vanishes at the compactness of a black hole ($m/R=0.5$) regardless
of the EOS dependent quantity $y$~\cite{DamourNagar2009,
BinningtonPoisson2009}.

Normal matter EOS behave approximately as polytropes for large
compactness.  However, for smaller compactness, the softer crust
becomes a greater fraction of the star, so the star is more
centrally condensed and $k_2$ smaller.  For strange quark matter, the EOS is extremely stiff near the minimum
density, and the star behaves approximately as an $n=0$ polytrope for
small compactness.  As the central density and compactness increase,
the softer part of the EOS has a larger effect, and the
star becomes more centrally condensed.

\begin{figure}[!htb]
\begin{center}
\includegraphics[width=70mm]{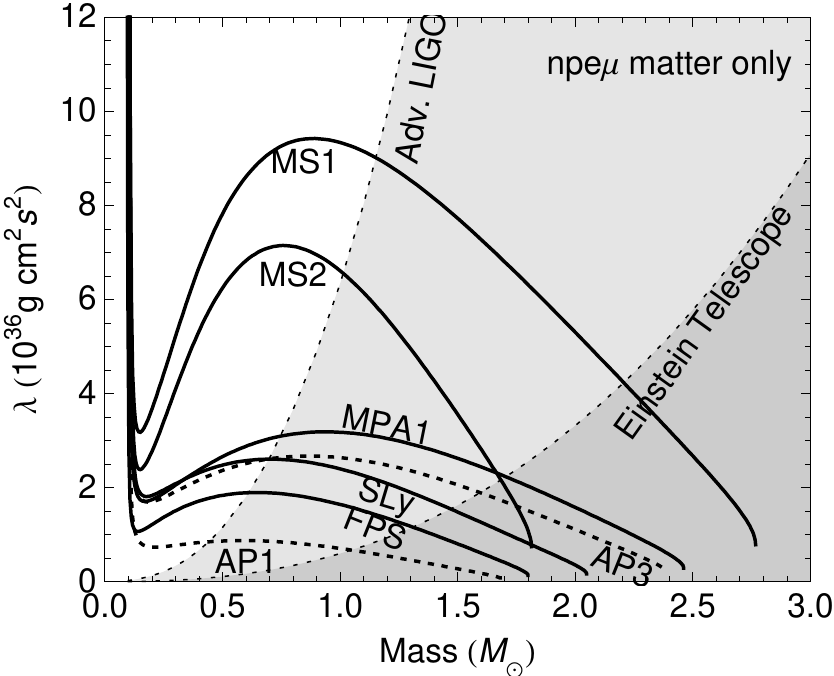}\\
\includegraphics[width=70mm]{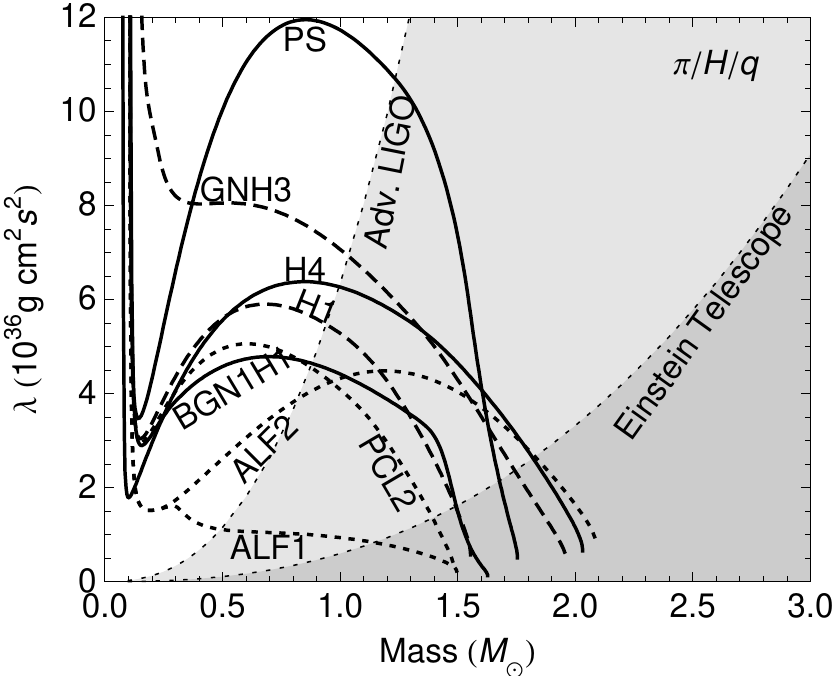}\\
\includegraphics[width=70mm]{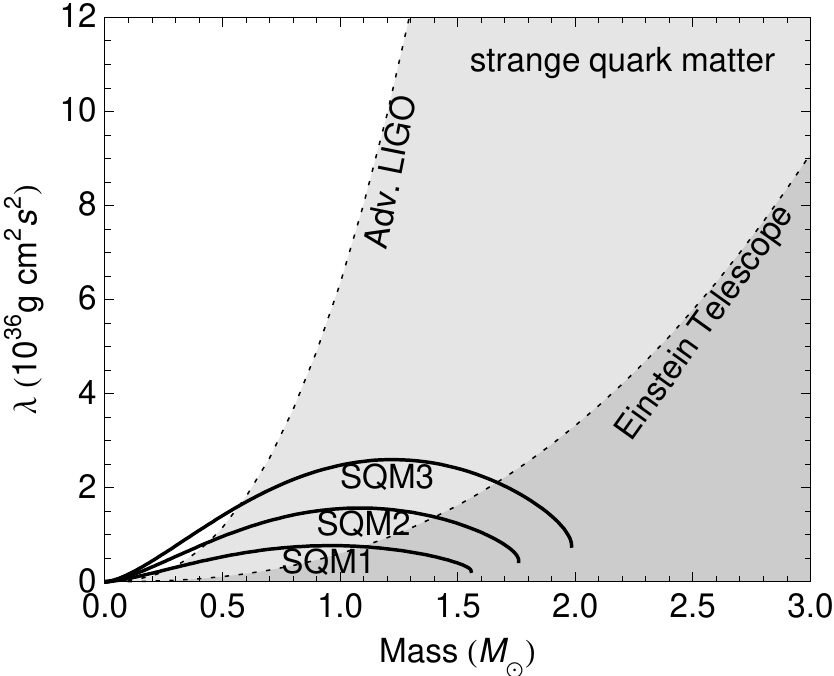}
\end{center}
\caption{ \label{fig:lambdaofm} Tidal deformability $\lambda$ of a
  single neutron star as a function of neutron-star mass for a range
  of realistic EOS.  The top figure shows EOS that only include
  $npe\mu$ matter; the middle figure shows EOS that also incorporate
  $\pi$/hyperon/quark matter; the bottom figure shows strange quark
  matter EOS.  The dashed lines between the various shaded regions represent the expected uncertainties in measuring
  $\lambda$ for an equal-mass binary inspiral at a distance of $D = 100$ Mpc as it passes through the gravitational wave frequency range 10~Hz--450~Hz.
  Observations with Advanced LIGO will be sensitive to $\lambda$ in
  the unshaded region, while the Einstein Telescope will be able to
  measure $\lambda$ in the unshaded and light shaded regions. See text
  below.}
\end{figure}

The parameter that is directly measurable by gravitational wave
observations of a binary neutron star inspiral is proportional to the
tidal deformability $\lambda$, which is shown for each candidate EOS
in Fig.~\ref{fig:lambdaofm}.  The values of $\lambda$ for the candidate EOS show a
much wider range of behaviors than for $k_2$ because $\lambda$ is
proportional to $k_2R^5$, and the candidate EOS produce a wide range
of radii (9.4--15.5~km for a 1.4~$M_\odot$ star for normal EOS and
8.9--10.9~km for the SQM EOS).  See Table~\ref{tab:lovenumber}.

\begin{table}[h]
  \caption{Properties of a 1.4~$M_\odot$ neutron star for the 18 EOS discussed in the text.}
\label{tab:lovenumber}
\begin{center}
\begin{tabular}{lcccc}
\hline\hline
EOS & $R$(km) & $m/R$ & $k_2$ & $\lambda(10^{36}$\,g\,cm$^2$\,s$^2)$\\
\hline
SLY & 11.74 & 0.176 & 0.0763 & 1.70\\
AP1 & 9.36 & 0.221 & 0.0512 & 0.368\\
AP3 & 12.09 & 0.171 & 0.0858 & 2.22\\
FPS & 10.85 & 0.191 & 0.0663 & 1.00\\
MPA1 & 12.47 & 0.166 & 0.0924 & 2.79\\
MS1 & 14.92 & 0.139 & 0.110 & 8.15\\
MS2 & 13.71 & 0.151 & 0.0883 & 4.28\\
\hline
PS & 15.47 & 0.134  & 0.104 & 9.19\\
BGN1H1 & 12.90 & 0.160 & 0.0868 & 3.10\\
GNH3 & 14.20 & 0.146 & 0.0867 & 5.01\\
H1 & 12.86 & 0.161 & 0.0738 & 2.59\\
H4 & 13.76 & 0.150 & 0.104 & 5.13\\
PCL2 & 11.76 & 0.176 & 0.0577 & 1.30\\
ALF1 & 9.90 & 0.209 & 0.0541 & 0.513\\
ALF2 & 13.19 & 0.157 & 0.107 & 4.28\\
\hline
SQM1 & 8.86 & 0.233 & 0.098 & 0.536\\
SQM2 & 10.03 & 0.206 & 0.136 & 1.38\\
SQM3 & 10.87 & 0.190 & 0.166 & 2.52\\
\hline \hline
\end{tabular}
\end{center}
\end{table}

For normal matter, $\lambda$ becomes large for stars near the
minimum mass configuration at roughly $0.1~M_\odot$ because
they have a large
radius. For masses in the expected mass range for binary inspirals,
there are several differences between EOS with only $npe\mu$ matter
and those with condensates. EOS with condensates have, on average, a
larger $\lambda$, primarily because they have, on average, larger
radii. The quark hybrid EOS ALF1 with a small radius (9.9~km for a
$1.4~M_\odot$ star) and the nuclear matter only EOSs MS1 and MS2
with large radii (14.9~km and 14.5~km, respectively, at
$1.4~M_\odot$) are exceptions to this trend.

For strange quark matter stars, there is no minimum mass, so the radius
(and therefore $\lambda$) approaches zero as the mass approaches zero.
At larger masses, the tidal deformability of SQM stars remains smaller
than most normal matter stars because, despite having large Love
numbers, the radii of SQM stars are typically smaller.

Error estimates $\Delta\lambda$ for an equal-mass
binary inspiral at 100~Mpc are also shown in
Fig.~\ref{fig:lambdaofm} for both Advanced LIGO and the Einstein
Telescope.  They will be discussed in the next section.

\section{Measuring effects on gravitational radiation}
\label{sec:radiation}
We wish to
calculate the contribution from realistic tidal effects to the phase
evolution and resulting gravitational wave spectrum of an
inspiraling neutron star binary. In the secular limit, where the
orbital period is much shorter than the gravitational radiation
reaction timescale, we consider the tidal contribution to the energy
$E$ and energy flux $dE/dt$
for a quasi-circular inspiral using the formalism developed
by Flanagan and
Hinderer~\cite{FlanaganHinderer2008}, which adds the following
leading-order terms to the post-Newtonian point-particle corrections
(PN-PP corr.):
\begin{eqnarray}
E(x) &=& -\frac{1}{2} M  \eta x  \biggl[1 + \text{(PN-PP corr.)}\nonumber\\
&& ~~~~~~~~ - 9 \frac{m_2}{m_1} \frac{\lambda_1} {M^5} x^5 + 1 \leftrightarrow 2 \biggr] \, , \\
\dot E (x) &=& -\frac{32}{5} \eta^2 x^{5} \biggl[1 + \text{(PN-PP corr.)}\nonumber\\
&& ~~~~~~~~ + 6 \frac{m_1+3m_2}{m_1} \frac{\lambda_1}{M^5} x^5 + 1 \leftrightarrow 2 \biggr].
\end{eqnarray}
Here $\lambda_1 = \lambda(m_1)$ and $\lambda_2 = \lambda(m_2)$ are the tidal deformabilities of stars 1 and 2, respectively.  $M=m_1+ m_2$ is the total mass, $\eta = m_1 m_2 / M^2$ is the dimensionless reduced mass,
and $x$ is the post-Newtonian dimensionless parameter given by $x=(\omega M)^{2/3}$, where $\omega$ is
the orbital angular frequency.  One can then use
\begin{equation}
dx/dt = \frac{\dot E}{dE/dx}
\end{equation}
to estimate the evolution of the quadrupole gravitational wave phase
$\Phi$ via $d\Phi/dt = 2\omega = 2x^{3/2}/M$.

\begin{figure}[tb]
\begin{center}
\includegraphics[width=70mm]{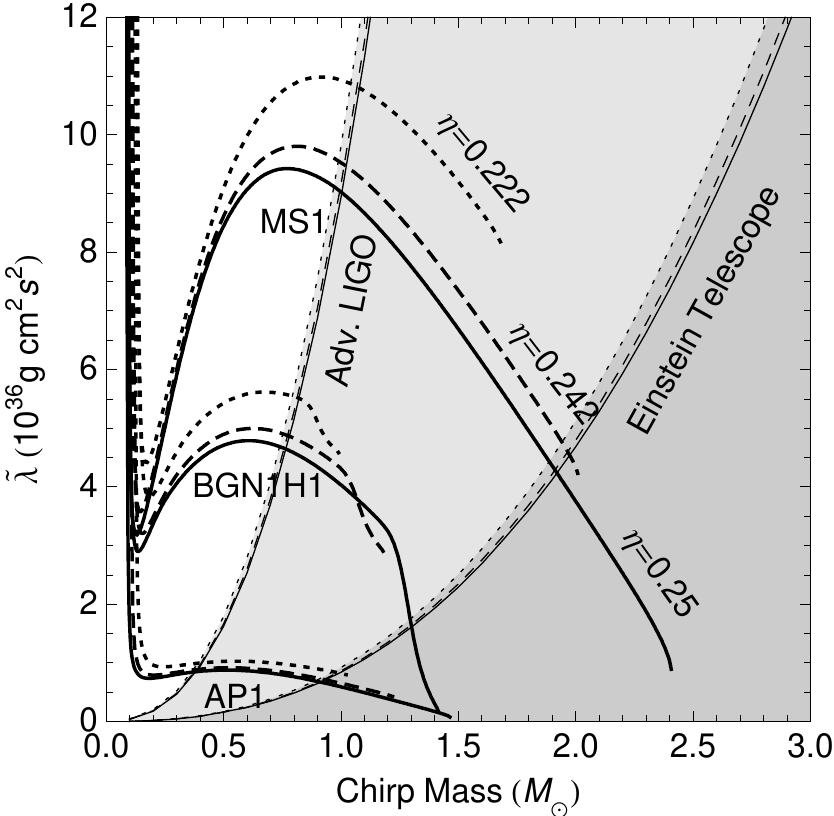}
\end{center}
\caption{ \label{fig:lambdatilde} Weighted $\tilde{\lambda}$ for a
  range of chirp mass $\mathcal{M}$ and dimensionless reduced mass
  $\eta$, for three of the EOSs considered above.  The values of
  $\eta$ equal to \{0.25, 0.242, 0.222\} correspond to the mass ratios
  $m_2/m_1 = $ \{1.0, 0.7, 0.5\}.
  Also plotted (as in Fig.~\ref{fig:lambdaofm}) are the uncertainties $\Delta\tilde\lambda$ in measuring $\tilde{\lambda}$ for a binary at 100 Mpc between 10~Hz--450~Hz.  The solid, dashed, and dotted curves correspond to $\Delta\tilde\lambda$ for $\eta= 0.25$, 0.242, and 0.222 respectively.}
\end{figure}

Each equation of state gives in this approximation a known phase
contribution as a function of $m_1$ and $m_2$, or as a function of the
total mass $M=m_1+m_2$ and the mass ratio $m_2/m_1$, via
$\lambda(m_1)$ and $\lambda(m_2)$ for that EOS. Although we
calculated $\lambda$ for individual neutron stars, the universality
of the neutron star core equation of state allows us to predict the
tidal phase contribution for a given binary system from each EOS.
Following \cite{FlanaganHinderer2008}, we discuss the constraint on
the weighted average
\begin{equation}
\tilde\lambda=\frac{1}{26}\left[\frac{m_1 + 12 m_2}{m_1}\lambda(m_1)
+ \frac{m_2 + 12 m_1}{m_2}\lambda(m_2)\right] \, ,
\end{equation}
which reduces to $\lambda$ in the equal mass case.  The contribution to
$d\Phi/dx$ from the tidal deformation, which adds linearly to the
known PP phase evolution, is
\begin{equation}
\left.\frac{d\Phi}{dx}\right|_{\text{T}} = -\frac{195}{8} \frac{
x^{3/2} \tilde \lambda}{M^5 \eta}.\label{eq:tidalterm}
\end{equation}
The weighted average $\tilde\lambda$ is plotted as a function of
chirp mass $\mathcal{M} = (m_1m_2)^{3/5}/M^{1/5}$ in
Fig.~\ref{fig:lambdatilde} for three of the EOS and for three values
of $\eta$: equal mass ($\eta=0.25$), large but plausible mass
ratio~\cite{BulikGondek2004} ($\eta=0.242$), and extremely large mass
ratio ($\eta=0.222$).

\begin{figure}[!bth]
\begin{center}
\includegraphics[width=70mm]{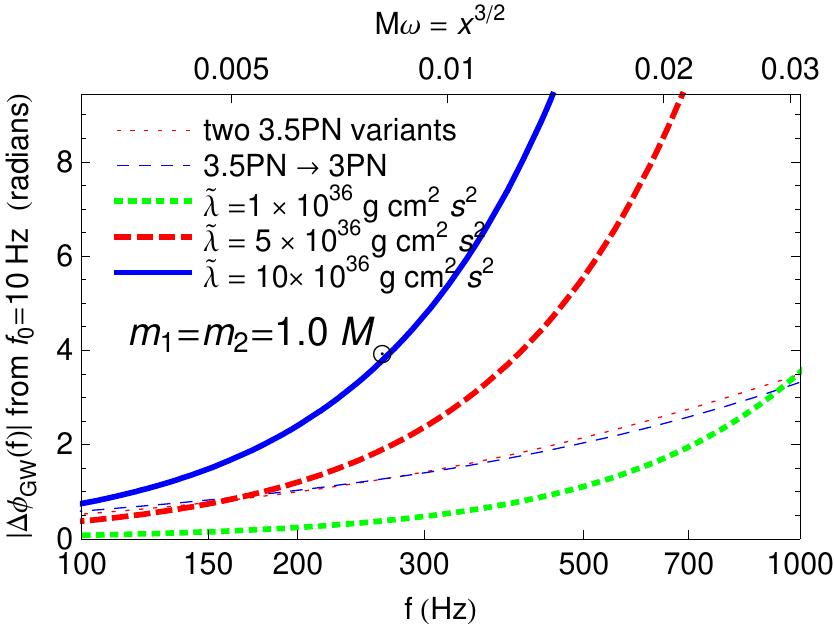}
\includegraphics[width=70mm]{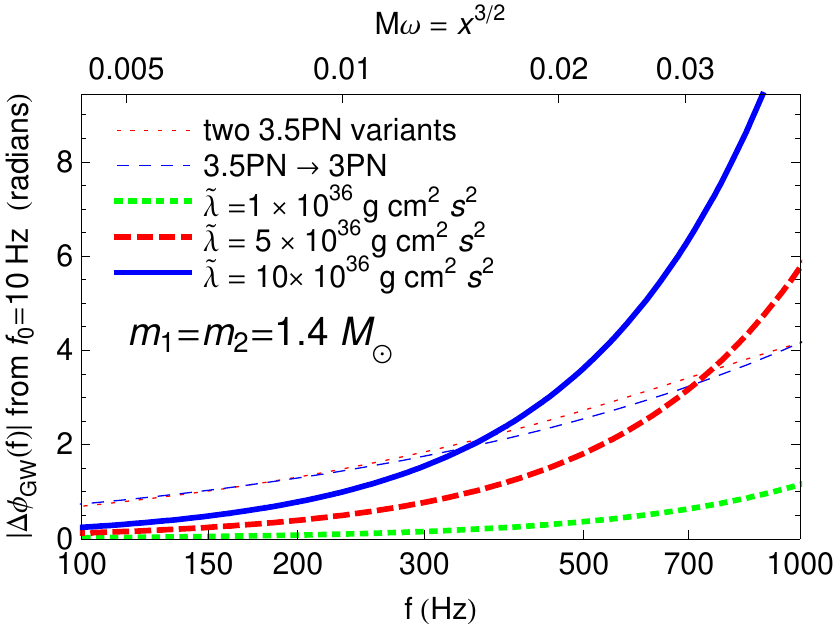}\\
\includegraphics[width=70mm]{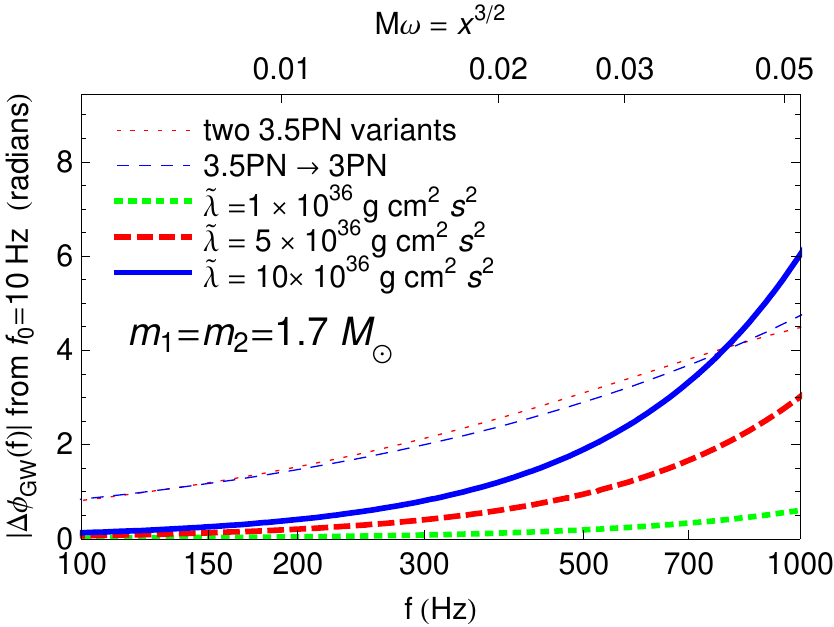}
\end{center}
\caption{ \label{fig:phases} The reduction in accumulated
  gravitational wave phase due to tidal effects, $\Phi_{3.5,PP}(f_{GW}) -
	\Phi_{3.5,\lambda}(f_{GW})$, is plotted with thick lines as a function of
	gravitational wave frequency, for a range of $\lambda$
	appropriate for realistic neutron star EOS and the masses considered.
	The 3.5 post-Newtonian TaylorT4 PN specification is used as the
	point-particle reference for the phase calculations. For reference, the
	difference in accumulated phase between 3.0 and 3.5 post-Newtonian orders
	of T4 (thin dashed line), and the difference between 3.5 post-Newtonian
	T4 and 3.5 post-Newtonian T1 (thin dotted line) are also shown. Phase
	accumulations are integrated from a starting frequency of 10 Hz.}
	\end{figure}

We can determine the significance of the tidal effect on
gravitational waveforms in a given frequency range by considering
the resulting change in phase accumulated as a function of
frequency. In the case of template-based searches, for example, a
drift in phase of half a cycle leads to destructive interference
between the signal and template, halting the accumulation of signal
to noise ratio. The phase contributions to binary neutron stars of
various masses from a range of realistic tidal deformabilities are
plotted in Fig.~\ref{fig:phases}.

The post-Newtonian formalism itself is sensitive to high-order
corrections at the frequencies at which the tidal effect becomes
significant; as reference, we show in Fig.~\ref{fig:phases} the phase
difference between the 3.0PN and 3.5PN expansions, as well
as that from varying the form of the post-Newtonian Taylor expansion
from T4 to T1.\footnote{For an explanation of the differences between
  T4 and T1, see~ \cite{DamourIyerSathyaprakash2001,
    BoyleBrownKidderEt2007}.}  An accurate knowledge of the underlying
point-particle dynamics will be important to resolve the effects of
tidal deformation on the gravitational wave phase evolution at these
frequencies.

The half-cycle or more contribution to the
gravitational wave phase at relatively low frequencies suggests that
this effect could be measurable.
Flanagan and Hinderer \cite{FlanaganHinderer2008} first calculated the
measurability for frequencies below 400~Hz, where the approximations
leading to the tidal phase correction are well-justified. We extend the
same computation of measurability to a range of masses and mass ratios. We
take noise curves from the projected NS-NS optimized Advanced LIGO
configuration~\cite{LIGOnoise}, as well as a proposed noise spectrum of the
Einstein Telescope~\cite{0810.0604}.  These noise curves are representative
of the anticipated sensitivities of the two detectors.
Our results do not change significantly for alternate configurations which
have similar sensitivities in the frequency range of interest.

We also extend the computation to a slightly higher cutoff
frequency. As estimated in the Appendix, our calculation should
still be fairly robust at 450~Hz, as the contributions to the phase
evolution from various higher order effects are $O( 10\%)$ of the
leading order tidal contribution. The uncertainty in the phase
contribution from a given EOS is therefore significantly smaller
than the order of magnitude range of phase contributions over the
full set of realistic EOS.

The rms uncertainty $\Delta \tilde \lambda$ in the measurement of $\tilde\lambda$ is computed using the
standard Fisher matrix formalism \cite{PoissonWill}.  Assuming a strong signal $h$ and Gaussian detector
noise, the signal parameters $\theta^i$ have probability distribution $p\left(\theta^i \right) \propto {\rm
exp}\left(-(1/2) {\,}{\Gamma_{ij}\delta \theta^i \delta \theta^j}\right)$,
where $\delta \theta^i = \theta^i-{\hat \theta}^i$ is the difference between the parameters and their best-fit values $\hat{\theta}^i$ and $ \Gamma_{ij} = (
\partial h/\partial \theta^i \, , \, \partial h/\partial \theta^j )$ is the
Fisher information matrix.  The parentheses denote the inner product defined in \cite{PoissonWill}.  The rms measurement error in
${\theta}^i$ is given by a diagonal element of the inverse Fisher, or covariance, matrix: $\Delta \theta^i = \sqrt{(\Gamma^{-1})^{ii}}$.

Using the stationary phase approximation and neglecting post-Newtonian corrections
to the amplitude, the Fourier transform of the waveform for spinning
point masses is given by $\tilde h(f) = {\cal A} f^{-7/6}
{\rm exp}\left(i\Psi\right)$, where the point-mass contribution to
the phase $\Psi$ is given to 3.5 post-Newtonian order in Ref.
\cite{lrr}. The tidal term \be \delta\Psi^{\rm tidal}=-\frac{117
\tilde \lambda x^{5/2}}{8 \eta M^5}\ee
obtained from Eq.~(\ref{eq:Fourierphasecorr}) adds linearly to this,
yielding a phase model with 7 parameters ($t_c,\phi_c, {\cal
M},\eta,\beta,\sigma,{\tilde \lambda}$), where $\beta$ and $\sigma$
are spin parameters. We incorporate the maximum spin constraint for
the NSs by assuming a Gaussian prior for $\beta$ and $\sigma$ as in
\cite{PoissonWill}. The uncertainties computed will depend on the
choice of point-particle phase evolution, but we assume this to be
exactly the 3.5PN form for the current analysis.

The rms measurement uncertainty of $\tilde \lambda$,
along with the uncertainties in chirp mass $\mathcal{M}$ and
dimensionless reduced mass $\eta $, are given in
Table~\ref{tab:error} and plotted in Figs.~\ref{fig:lambdaofm}
and~\ref{fig:lambdatilde}, from a single-detector observation of a
binary at 100 Mpc distance with amplitude averaged over inclinations
and sky positions.
 If the best-fit $\tilde\lambda$ is
zero, this represents a 1-$\sigma$ upper bound on the physical
$\tilde\lambda$. A signal with best-fit $\tilde\lambda\geq \Delta
\tilde \lambda$ would allow a measurement rather than a constraint
of $\tilde \lambda$, with 1-$\sigma$ uncertainty of $\Delta \tilde
\lambda$.

We obtain the following approximate formula for the rms measurement
uncertainty $\Delta\tilde\lambda$, which is accurate to better than
$ 4\%$ for the range of masses $0.1\ M_\odot\leq m_1, m_2 \leq 3.0\
M_\odot$ and cutoff frequencies $400 {\rm\ Hz}\leq f_{\rm end}\leq
500{\rm\ Hz}$:
\begin{equation}
\Delta \tilde \lambda
\approx \alpha
\left(\frac{M}{M_\odot}\right)^{2.5}\left(\frac{m_2}{m_1}\right)^{0.1}\left(\frac{f_{\rm
end}}{\rm Hz}\right)^{-2.2}\left(\frac{D}{100 {\rm
Mpc}}\right),
\label{eq:fittingligo}
\end{equation}
where $\alpha =
1.0\times 10^{42}$~g\,cm$^2$\,s$^2$ for a single Advanced LIGO detector and
$\alpha = 8.4\times 10^{40}$~g\,cm$^2$\,s$^2$ for a single Einstein
Telescope detector.

Our results show that the measurability of tidal effects
decreases steeply with the total mass of the binary.  Estimates of the
measurement uncertainty for an equal-mass binary inspiral in a single
detector with projected sensitivities of Advanced LIGO and the Einstein
Telescope, at a volume-averaged distance of 100 Mpc and using only the
portion of the signal between $10-450$ Hz, are shown in
Fig.~\ref{fig:lambdaofm}, together with the values of $\lambda$ predicted
by various EOS models. Measurability is less sensitive to mass ratio, as
seen in Fig.~\ref{fig:lambdatilde}.
Comparing the magnitude of the resulting upper
bounds on $\lambda$ with the expected range for realistic EOS, we
find that the predicted $\lambda$ are greatest and the
measurement uncertainty $\Delta\lambda$ is
smallest for neutron stars at the low end of the expected mass range
for NS-NS inspirals of ($1\ M_\odot-1.7\ M_\odot$)~\cite{Stairs}.

In a single Advanced LIGO detector, only extremely stiff EOS could be
constrained with a typical 100 Mpc observation. However, a rare nearby
event could allow more interesting constraints, as the uncertainty scales
as the distance to the source.  Rate estimates for detection of binary
neutron stars are often given in terms of a minimum signal-to-noise
$\rho_c=8$; a recent estimate \cite{CoalRates} is between 2 and 64 binary
neutron star detections per year for a single Advanced LIGO
interferometer with a volume averaged range of 187 Mpc. The rate of
binaries with a volume averaged distance smaller than 100 Mpc translates to
roughly $\left(100/187\right)^3 \simeq 15 \%$ of this total detection rate,
but over multiple years of observation a rare event could give measurements
of $\tilde{\lambda}$ with uncertainties smaller than the values in
Table~\ref{tab:error} (e.g.\ with half the tabled uncertainty at  $1.9\%$
the total NS-NS rate).

Using information from a network of $N$ detectors with
the same sensitivity decreases the measurement uncertainty by approximately
a factor of $1/\sqrt{N}$~\cite{CutlerFlanagan1994}, giving more reason for optimism.  However, we should also note that, in some ways, our estimates of uncertainty are already too optimistic.  First, $\Delta \lambda$ only represents a $68\%$ confidence in the measurement; a $2 \Delta
\lambda$ error bar would give a more reasonable $95\%$ confidence.  In addition, our Fisher matrix estimates are likely to somewhat
underestimate the measurement uncertainty in real non-Gaussian noise.

In contrast to Advanced LIGO, an Einstein Telescope detector with currently
projected noise would be sensitive to tidal effects for typical binaries,
using only the signal below 450~Hz at 100~Mpc. The tidal signal in this
regime would provide a clean signature of the neutron star core equation of
state.  However, an accurate understanding of the underlying point-particle
phase evolution is still important to confidently distinguish EOS effects.

\begin{table}[h]
\caption{The rms measurement error in various binary parameters
(chirp mass ${\cal M}$, dimensionless reduced mass $\eta$, and
weighted average $\tilde \lambda$ of the tidal deformabilities) for
a range of total mass $M$ and mass ratio $m_2/m_1$, together with
the signal to noise ratio $\rho$, using only the information in the
portion of the inspiral signal between $10{\rm\ Hz}\leq f \leq 450 {\rm\ Hz}$.
The distance is set at 100~Mpc, and the amplitude
is averaged over sky position and relative inclination.}
\label{tab:error}

\begin{center}
Advanced LIGO \\
\begin{tabular}{cccccc}

\hline\hline M $(M_\odot)$ &$m_2/m_1$  & $\Delta \cal M / \cal M$ &
$\Delta \eta/ \eta$ & $\Delta \tilde
\lambda(10^{36}$\,g\,cm$^2$\,s$^2)$ & $\rho$
\\\hline
2.0 & 1.0& 0.00028 & 0.073 & 8.4 & 27\\
%
%
2.8 & 1.0 & 0.00037 & 0.055 & 19.3 & 35 \\
%
%
3.4 & 1.0 & 0.00046 & 0.047 & 31.3 & 41\\
%
2.0 & 0.7 & 0.00026 & 0.058 & 8.2 & 26\\
2.8 & 0.7 & 0.00027 & 0.058 & 18.9 & 35\\
3.4 & 0.7 & 0.00028 & 0.055 & 30.5 & 41\\
2.8 & 0.5 & 0.00037 & 0.06 & 17.8 & 33\\
\hline\hline
\end{tabular}
\\
$~$\\
Einstein Telescope \\
\begin{tabular}{cccccc}
\hline\hline M $(M_\odot)$ &$m_2/m_1$  & $\Delta \cal M / \cal M$ &
$\Delta \eta / \eta$ & $\Delta \tilde
\lambda(10^{36}$\,g\,cm$^2$\,s$^2)$ & $\rho$
\\\hline
2.0 & 1.0 & 0.000015 & 0.0058 & 0.70 & 354\\
%
%
2.8 & 1.0 & 0.000021 & 0.0043 & 1.60 & 469\\
%
%
3.4 & 1.0 & 0.000025 & 0.0038 & 2.58 & 552\\
2.0 & 0.7 & 0.000015 & 0.0058 & 0.68 & 349\\
2.8 & 0.7 & 0.000021 & 0.0045 & 1.56 & 462\\
3.4 & 0.7 & 0.000025 & 0.0038 & 2.52 & 543\\
2.8 & 0.5 & 0.000020 & 0.0048 & 1.46 & 442\\
\hline\hline
\end{tabular}
\end{center}
\end{table}

Expected measurement uncertainty will decrease if we can extend the
calculation later into the inspiral.  From
Eq.~\eqref{eq:fittingligo}, $\Delta\tilde\lambda$ at 500~Hz is
approximately 79\% of its value at 450~Hz. The dominant source of
error in the tidal phasing at these frequencies are post-Newtonian
effects which scale as $\lambda x^{7/2}$ and do not depend on any
additional EOS parameters.  These terms are computed in
Ref.~\cite{Vinesflanagan}, and when they are incorporated into the
analysis, the resulting phase evolution model can be used at
slightly higher frequencies.  These terms also add $\sim 10\%(f/450\
\rm Hz)^{2/3}$ to the strength of the tidal signature.

Higher-order tidal effects and nonlinear hydrodynamic couplings,
which depend on unknown NS microphysics, are smaller than post-Newtonian effects by
factors of $\sim x$ and $\sim x^2$, so they become important later in the
inspiral, where the adiabatic approximation that the mode frequency
is large compared to the orbital frequency also breaks down.   At
this point we can no longer measure only $\tilde\lambda$, but an EOS-
dependent combination of effects including higher multipoles,
nonlinearity, and tidal resonances.

However, information in the late inspiral could also constrain the
underlying neutron-star EOS. Read et al.~\cite{ReadMarkakisShibataEt2009}
estimated potential measurability of EOS effects in the last few orbits of
binary inspiral, where the gravitational
wave frequency is above 500~Hz, using full numerical simulations.  The EOS used for the simulation was
systematically varied by shifting the pressure in the core while keeping
the crust fixed. The resulting models were
parameterized, either by a fiducial pressure or by the radius of the isolated NS
model, and measurability in Advanced LIGO was estimated.  Such numerical simulations
include all the higher order EOS effects described above, but the $l=2$ tidal
deformability parameter $\lambda$ should remain the dominant source of
EOS-dependent modification of the phase evolution. We therefore expect it
to be a better choice for a single parameter to characterize EOS effects on
the late inspiral.

The numerically simulated models of~\cite{ReadMarkakisShibataEt2009} can be
re-parameterized by the $\lambda$ of the 1.35 $M_\odot$ neutron stars
considered\footnote{The piecewise polytrope EOS \{2H, H, HB, B, 2B\} have
$\lambda_{1.35 M_\odot}$ of \{0.588, 1.343, 1.964, 2.828, 10.842\}$\times
10^{36}$\,g\,cm$^2$\,s$^2$, respectively.}. The uncertainty of measurement
for the new parameter $\lambda$ can be estimated from Tables II-V of
\cite{ReadMarkakisShibataEt2009}. In the broadband Advanced LIGO
configuration of Table IV, it is between 0.3 and 4$\times
10^{36}$\,g\,cm$^2$\,s$^2$ for an optimally oriented 100 Mpc binary, or
between 0.7 and 9$\times 10^{36}$\,g\,cm$^2$\,s$^2$ averaged over sky
position and orientation. However, in the NS-NS optimized LIGO
configuration of Table III, which is most similar to the Advanced LIGO
configuration considered in this paper, the expected measurement
uncertainty is more than several times $\lambda$ for all models. These estimates
should be considered order-of-magnitude, as numerical simulation errors are
significant, and the discrete sampling of a parameter space allows only a
coarse measurability estimate which neglects parameter correlations. In
contrast to the perturbative/post-Newtonian estimate of EOS effects
calculated in this paper, EOS information in the signal before the start of
numerical simulations is neglected. The estimate is complementary to the
measurability below 450~Hz estimated in this paper.

\section{Conclusion}

We have calculated the relativistic $l=2$ Love number $k_2$ and resulting tidal deformability
$\lambda$ for a wide range of realistic EOS in addition to polytropes.  These EOS
have tidal deformabilities that differ by up to an order of magnitude in
the mass range relevant for binary neutron stars.
However, the estimated uncertainty
$\Delta\tilde\lambda$ for a binary neutron star inspiral at 100~Mpc using
the Advanced LIGO sensitivity below 450~Hz is greater than the largest
values of $\tilde\lambda$ except for very low-mass binaries.  The
uncertainty for the Einstein Telescope, on the other hand, is approximately
an order of magnitude smaller than for Advanced LIGO, and a measurement of
$\tilde\lambda$ will rule out a significant fraction of the EOS.

Advanced LIGO can place a
constraint on the space of possible EOS by obtaining a
 $95 \%$ confidence upper limit of $\tilde\lambda(\mathcal{M}, \eta)
 \lesssim 2 \Delta\tilde\lambda(\mathcal{M}, \eta)$.  The tables in
 Sec.~\ref{sec:radiation} can also be scaled as follows: For a network of
 $N$ detectors the uncertainty scales roughly as
 $\Delta\tilde\lambda/\sqrt{N}$, and for a closer signal we have
 $\Delta\tilde\lambda (D/100\textrm{ Mpc})$.

\acknowledgments

We thank S. Hughes, E. Flanagan, and J. Friedman for helpful suggestions and
J. Creighton for carefully reading the manuscript. The work
was supported in part by NSF Grant PHY-0503366, and by the Deutsche
Forschungsgemeinschaft SFB/TR7.  TH gratefully acknowledges support from
the Sherman Fairchild postdoctoral fellowship, and BL also thanks the
Wisconsin Space Grant Consortium fellowship program for support.  RNL was
supported by NSF Grant PHY-0449884 and the NASA Postdoctoral Program,
administered by Oak Ridge Associated Universities through a contract with
NASA.

\appendix*

\section{Accuracy of the phasing model}
\label{app:errors}

To assess the accuracy of the simple phase evolution
model, we compute the corrections to the tidal phase perturbation
due to several EOS-dependent effects: the leading order finite
mode-frequency terms, higher order tidal effects, and nonlinear
hydrodynamic couplings. For simplicity, we will only derive the
phase corrections for one star with internal degrees of freedom
coupled to a point mass. The terms for the other star simply add.
 For such a binary system, the Lagrangian can then be written
as
\begin{eqnarray}
L &=& {1 \over 2} \eta M {\dot r}^2 + {1 \over 2} \eta M r^2 {\dot\varphi}^2 + { \eta M^2 \over r}\nonumber\\
  & & -  {1 \over 2}Q_{ij} {\cal E}_{ij}  + {1 \over 4 \lambda \omega_0^2} \left({\dot Q}_{ij} {\dot Q}_{ij} - \omega_0^2 Q_{ij}Q_{ij}\right)\nonumber\\
  & & - \frac{1}{6}Q_{ijk}{\cal E}_{ijk}+\frac{1}{12\lambda_3\omega_{03}^2}\left(\dot Q_{ijk}\dot Q_{ijk} - \omega_{03}^2 Q_{ijk}Q_{ijk}\right)\nonumber\\
  & & - \frac{\alpha}{\lambda^3}Q_{ij} Q_{jk}Q_{ki}.
\end{eqnarray}
Here, the star's static mass quadrupole $Q_{ij}$ parameterizes the
$l=2$ modes of the star, which can be treated as harmonic
oscillators that are driven below their resonant frequency by the
companion's tidal field. The tensor $Q_{ijk}$ parameterizes the
star's mass octupole degrees of freedom, and ${\cal E}_{ij}$ and
${\cal E}_{ijk}$ are the $l=2$ and $l=3$ tidal tensors respectively,
which are given by ${\cal E}_{ij}=\partial_i\partial_j (-m_2/r)$ and
 ${\cal E}_{ijk}=\partial_i\partial_j\partial_k (-m_2/r)$ in Newtonian gravity.
The $l=3$ deformability constant $\lambda_3$ is defined by $Q_{ijk}
= -\lambda_3 \mathcal{E}_{ijk}$. The quantities $\omega_0$ and
$\omega_{03}$ are the $l=2$ and $l=3$ $f$-mode frequencies, and $\alpha$ is
a coupling constant for the leading order nonlinear hydrodynamic
interactions. In general, one would need to sum over the contributions from
all the modes, but other modes contribute negligibly in the regime of
interest for the above model (see~\cite{FlanaganHinderer2007}).
Post-Newtonian effects on the Lagrangian for the binary are derived in Ref.
\cite{Vinesflanagan} and can simply be added to those derived here.

We will be interested in finding an effective description of the
dynamics of the system for quasi-circular inspirals in the adiabatic
limit, where the radiation reaction timescale is long compared to
the orbital timescale. From equilibrium solutions to the
Euler-Lagrange equations derived from this Lagrangian, the following
radius-frequency relation is obtained:
\begin{widetext}
\begin{equation}
r(\omega) = M^{1/3}\omega ^{-2/3} \left[1
                + \frac{3 \lambda m_2\omega ^{10/3}}{M^{5/3}m_1}
                + \frac{9 \lambda m_2 \omega^{10/3}}{M^{5/3} m_1} \frac{\omega^2}{\omega_0^2}
                + \frac{20 \lambda_3 m_2 \omega^{14/3}}{M^{7/3} m_1}
                - \frac{18 \alpha m_2^2 \omega ^{16/3}}{M^{8/3} m_1}
                - \frac{27 \lambda^2 m_2^2 \omega^{20/3}}{M^{10/3}m_1^2}\right],
\end{equation}
The equilibrium energy, obtained by reversing the signs of the
potential energy terms in the Lagrangian, is given by:
\begin{equation}
E = -\frac{1}{2} \eta M^{5/3} \omega^{2/3} \left[1
   - \frac{9 \lambda m_2\omega^{10/3}}{M^{5/3} m_1}
   - \frac{45 \lambda m_2\omega ^{10/3}}{M^{5/3} m_1} \frac{\omega^2}{\omega_0^2}
   - \frac{65 \lambda_3 m_2 \omega ^{14/3}}{M^{7/3} m_1}
   + \frac{60 \alpha m_2^2 \omega^{16/3}}{M^{8/3} m_1}
   + \frac{63 \lambda^2 m_2^2\omega^{20/3}}{M^{10/3}m_1^2}\right].
\end{equation}
The correction to the energy flux $\dot E = - \tfrac{1}{5} \langle \dddot Q_{ij}^T \dddot Q_{ij}^T \rangle$, where $Q_{ij}^T = \mu r^2(n^i n^j - \tfrac{1}{3}\delta_{ij}) + Q_{ij}$ is the total quadrupole moment, is
\begin{eqnarray}
\dot E &=& -\frac{32}{5} \eta^2 M^{10/3} \omega^{10/3}  \left[1
            + \frac{6 \lambda  \omega ^{10/3}}{M^{2/3} m_1}\left(2\frac{m_2}{M}+1\right)
            + \frac{12 \lambda  \omega ^{10/3}}{M^{2/3} m_1}\frac{\omega^2}{
            \omega_0^2}\left(3\frac{m_2}{M}+2\right)+ \frac{80 \lambda_3 m_2 \omega^{14/3}}{M^{7/3}m_1}\right.\nonumber\\
           &&\left. ~ ~ ~ ~ ~ ~ ~ ~ ~ ~ ~ ~ ~~ ~
            - \frac{36 \alpha m_2 \omega^{16/3}}{M^{5/3} m_1}\left(2\frac{m_2}{M}+1\right)
            +\frac{9\lambda^2\omega^{20/3}}{M^{4/3}m_1^2}\left(1-\frac{2m_2}{M}-\frac{6m_2^2}{M^2}\right) \right].
\end{eqnarray}
Using the formula $d^2\Psi/d\omega^2=2 (dE/d\omega)/\dot E$ in the
stationary phase approximation and integrating twice leads to the
final expression for the tidal phase correction:
\begin{eqnarray}
\delta\Psi &=& - \frac{9 \lambda x^{5/2}}{16 \eta
M^5}\left(\frac{m_1 + 12m_2}{m_1}\right) - \frac{45 \lambda
x^{5/2}}{1408 \eta M^5}\frac{\omega^2}{\omega_0^2}\left(\frac{8m_1 +
155m_2}{m_1}\right) - \frac{125}{12}\frac{\lambda_3 x^{9/2}}{\eta
M^7}\frac{m_2}{m_1} \nonumber\\
&&+ \frac{135 \alpha m_2 x^{11/2}}{352 \eta M^8}\left(\frac{m_1 +
18m_2}{m_1}\right)-\frac{3\lambda^2x^5}{64\eta
M^{10}}\left(\frac{M^2-2m_2 M-83m_2^2}{m_1^2}\right).
\label{eq:Fourierphasecorr}
\end{eqnarray}
\end{widetext}

 We will analyze the information contained in the
portion of the signal at frequencies $f \leq 450 \, {\rm Hz}$. This
is slightly higher than previously considered, and we now argue that
in this frequency band, the simple model of the phase correction is
still sufficiently accurate for our purposes. We will evaluate all
of the corrections for the case of equal masses $m_1=m_2\equiv m$.
An estimate of the fractional errors for the case of $m=1.4\
M_\odot$ and $R=15$~km is given in parentheses.

\begin{enumerate}

\item {\it Post-1-Newtonian corrections} ($\sim 10\%$).\\
These corrections give rise to terms $\propto \lambda x^{7/2}$ that add
to those in Eq.~(\ref{eq:Fourierphasecorr}). The explicit form of these
terms is computed in Ref.~\cite{Vinesflanagan} and they depend on
the NS physics only via the same parameter $\lambda$ as the
Newtonian tidal terms, so they can easily be incorporated into the
data analysis method. Preliminary estimates indicate that for equal
masses, these post-1 Newtonian effects will increase the tidal
signal.

\item {\it Adiabatic approximations} ( $\alt 1\%$).\\
The approximation that the radiation reaction time is much longer
than the orbital time is extremely accurate, to better than $1\%$;
see Fig.~2 of Ref.~\cite{FlanaganHinderer2008}, which compares the
phase error obtained from numerically integrating the equations of
motion supplemented with the leading order gravitational wave
dissipation terms to that obtained analytically using
the adiabatic approximation. \\
The accuracy of the approximation $\omega\ll\omega_0$ can be
estimated from the fractional correction
to~(\ref{eq:Fourierphasecorr}), which is $\sim
(815/1144)(\omega/\omega_0)^2 \sim 0.18 (f/f_0)^2$, where $f =
\omega / \pi$ and $f_0 = \omega_0 / (2 \pi)$. For typical NS models
the $l=2$ $f$-mode frequency is~\cite{1999LRRmodes}
\begin{equation}
\frac{f_0}{\rm kHz}\approx
0.78 + 1.64 \left(\frac{m}{1.4\ M_\odot}\right)^{1/2}\left(\frac{R}{10\ {\rm km}}\right)^{-3/2},
\end{equation}
so that the fractional correction is $\sim 0.012$ for $f = 450$~Hz
and for a conservatively low $f$-mode frequency of $f_0 = 1700$~Hz.

\item {\it Higher order tidal effects} ($\alt 0.7 \%$).\\
The $l=3$ correction to the gravitational wave phase
(\ref{eq:Fourierphasecorr}) is smaller than the $l=2$ contribution
by a factor of $\sim (25/351)(k_3/k_2)(m/R)^{-2} x^2 \sim 0.007$,
for $m/R=0.14$ and a stiff $n=0.5$ polytrope. Here, we have defined
the $l=3$ Love number $k_3=(15/2)\lambda_3 R^{-7}$ and used the
values $k_2=0.17$ and $k_3=0.06$ from
Ref.~\cite{BinningtonPoisson2009}.

\item {\it Nonlinear hydrodynamic corrections} ($\sim 0.1\%$).\\
The leading nonlinear hydrodynamic corrections are characterized by
the coupling coefficient $ \alpha/\lambda^3$ in the action. The size
of this parameter can be estimated by comparing the Newtonian $k_2$
to the coupling constants in Lai's ellipsoidal models (e.g.\ Table 1
of \cite{1993ApJS...88..205L}) to be $\omega^2\alpha/\lambda\sim
2\times 10^{-3}$. The nonlinear self-coupling term in
Eq.~(\ref{eq:Fourierphasecorr}) is smaller than the leading $l=2$
term by a factor $- 285 \alpha
 \omega^2/(572\lambda) \sim 0.001$.

\item {\it Spin corrections} ($\lesssim 0.3\%$).\\
Fractional corrections to the tidal signal due to spin scale as
\begin{equation}
\frac{\delta \Psi_{\rm spin}}{\delta \Psi_{\rm tidal}} \propto
\left(\frac{\omega_{\rm spin}}{\omega_{\rm max}}\right)^2,
\end{equation}
where $\omega_{\rm max}$ is the
maximum rotational frequency the star can have before breakup, which
for most NS models is $>2 \pi (1000\ {\rm Hz})$. The observed NS-NS
binaries which will merge within a Hubble time have spin periods of
$\sim 23-104$~ms, and near the coalescence they will have slowed down
due to e.g.\ magnetic braking, with final spin periods of $\sim
50-130$ms. The fractional corrections to the tidal signal due to the
spin are then $\lesssim 0.3\%$.

If the stars have spin, there will also be a spin-induced correction
to the phase. As discussed above, the slow-rotation limit is likely
to be the relevant regime for our purposes, and using similar
methods as for the tidal corrections leads to a phase correction
which scales as $\delta \Psi_{\rm s}\propto n_2 R^2/(\eta M^2
x^{1/2})\omega_{\rm spin}^2/(m_1/R^3)$, where $n_2$ is the
rotational Love number, which for Newtonian stars is the same as the
tidal Love number $k_2$ and $\omega_{\rm spin}$ the spin frequency.
The scaling of the spin term as $\propto x^{-1/2}$ shows that only
at large separation do spin effects dominate over tidal effects,
which scale as $\propto x^{5/2}$.

\item {\it Nonlinear response to the tidal field} ($\alt 3\%$).\\
We have linearized in $\lambda$.  Including terms $\propto\lambda^2$ gives a fractional
correction in Eq.~(\ref{eq:Fourierphasecorr}) of $- (83/7488) k_2
R^5 x^{5/2}/m^5 = -4.8\times 10^{-11}k_2 (m/M_\odot)^{-10/3} (R/{\rm
km})^5 (f/{\rm Hz})^{5/3} = -0.31 k_2$.

\item {\it Viscous dissipation} (negligible).\\
There have been several analytical and numerical
studies of the effect of viscosity during the early part of the
inspiral, e.g.\ \cite{Kochanek1992, BildstenCutler1992}. They
found that viscous dissipation is negligible during the early
inspiral if the volume-averaged shear viscosity $\eta_{\rm shear}$
is
\begin{equation}
\eta_{\rm shear}\lesssim 10^{29}\left(\frac{r}{R}\right)^2 {\rm
g~cm}^{-1}s^{-1}.
\end{equation}
The expected microscopic
viscosity of NSs is~\cite{CutlerLindblom}
\begin{equation}
\eta_{\rm micr} \sim 10^{22}\left(\frac{\rho}{10^{14}{\rm g
~cm}^{-3}}\right)^{9/4}\left(\frac{T}{10^{6}{\rm K}}\right)^{-2}
{\rm g~cm}^{-1}s^{-1},
\end{equation}
which is orders of magnitude too small to
lead to any significant effect. A variety of other likely sources of
viscosity, e.\ g.\ the breaking or crumpling of the crust, are also
insignificant \cite{BildstenCutler1992, Kochanek1992} in the regime
of interest to us.
\end{enumerate}

Thus, systematic errors in the measured value of $\lambda$ due to
errors in the model should be $O(10\%)$, which is small compared
to the current uncertainty of an order of magnitude in $\lambda$.

\bibliography{apsidal}
\end{document}